 \documentclass[twocolumn,prb,showpacs,amsmath,amssymb,floatfix]{revtex4}

\usepackage{graphicx}
\usepackage{dcolumn}
\usepackage{bm}
\usepackage{xspace}

\newcommand{\eg}{e.\,g.\@\xspace}   
\newcommand{\ie}{i.\,e.\@\xspace}   
\newcommand{\cf}{cf.\@\xspace}      
\newcommand{\tetal}{\emph{et.\@\xspace al.\@\xspace}}

\newcommand{\tref}{Ref.\@\xspace}   
\newcommand{\trefs}{Refs.\@\xspace} 

\newcommand{\teq}{Eq.\@\xspace}     
\newcommand{\teqs}{Eqs.\@\xspace}   

\newcommand{\tfig}{Fig.\@\xspace}   
\newcommand{\tfigs}{Figs.\@\xspace} 

\newcommand{\tchap}{Sect.\xspace}      
\newcommand{\tchapter}{Section\xspace} 

\newcommand{\lz}{\ell=0}
\newcommand{\li}{\ell=\infty}
\newcommand{\ldep}{(\ell)}
\newcommand{\crit}{\text{c}}
\newcommand{\diff}{\text{d}}
\newcommand{\eff}{\text{eff}}

\newcommand{\gc}{g_{\crit}}

\newcommand{\ga}{\gamma}
\newcommand{\gac}{\ga_\crit}

\newcommand{\om}{\omega}
\newcommand{\dom}{/\om}
\newcommand{\jdom}{J\dom}
\newcommand{\gdom}{g\dom}
\newcommand{\gcdom}{\gc\dom}

\newcommand{\al}{\alpha}
\newcommand{\ale}{\al_{\eff}}

\newcommand{\alc}{\al_{\crit}}

\newcommand{\etal}{\eta\ldep}

\newcommand{\bd}{b^{\dagger}}
\newcommand{\bi}{b_i}
\newcommand{\bdi}{b_i^{\dagger}}

\newcommand{\bhi}{\hat{b}_i}
\newcommand{\bhdi}{\hat{b}_i^{\dagger}}

\newcommand{\h}{{\mathcal{H}}}
\newcommand{\hl}{\h\ldep}
\newcommand{\hs}{\h_{\text{S}}}
\newcommand{\hb}{\h_{\text{B}}}
\newcommand{\hsb}{\h_{\text{SB}}}
\newcommand{\hd}{\h_{\diff}}
\newcommand{\he}{\h_{\text{eff}}}
\newcommand{\hsl}{\h_{\text{S}}\ldep}
\newcommand{\hbl}{\h_{\text{B}}\ldep}
\newcommand{\hsbl}{\h_{\text{SB}}\ldep}
\newcommand{\hdl}{\h_{\diff}\ldep}

\newcommand{\opa}{{\mathcal{A}}}
\newcommand{\opad}{\opa^{\dagger}}
\newcommand{\opadi}{\opad_i}
\newcommand{\opai}{\opa_i}

\newcommand{\opail}{\opai\ldep}
\newcommand{\opadil}{\opadi\ldep}

\newcommand{\opspure}{{\mathbf{S}}}
\newcommand{\opsgen}[1]{\opspure_{#1}}
\newcommand{\ops}[1]{\opsgen{i#1}}
\newcommand{\sprgen}[2]{\opsgen{#1}\opsgen{#2}}

\newcommand{\spr}[2]{\sprgen{i#1}{i#2}}

\newcommand{\dell}{\diff\ell}
\newcommand{\diffell}[1]{\frac{\diff#1}{\dell}}

\newcommand{\NN}{C}

\newcommand{\NNspin}{\langle\spr{}{+1}\rangle}
\newcommand{\NNNspin}{\langle\spr{}{+2}\rangle}

\newcommand{\limellinf}{\ell\rightarrow\infty}
\newcommand{\onehalf}{1/2}
\newcommand{\oh}{\onehalf}
\newcommand{\spinonehalf}{S\!=\!\oh}

\begin{document}
\setlength{\unitlength}{1mm}
\title{Spin-Phonon Chains with Bond Coupling}

\author{Carsten Raas}%
\email{e-mail: cr@thp.uni-koeln.de}%
\affiliation{Institut f\"ur Theoretische Physik, Universit\"at zu K\"oln, Z\"ulpicher Str. 77, D-50937 K\"oln, Germany.}

\author{Rainer W. K\"uhne}%
\affiliation{Institut f\"ur Theoretische Physik, Universit\"at Dortmund, D-44221 Dortmund, Germany.}%

\author{Ute L\"ow}%
\email{ul@thp.uni-koeln.de}%
\affiliation{Institut f\"ur Theoretische Physik, Universit\"at zu K\"oln, Z\"ulpicher Str. 77, D-50937 K\"oln, Germany.}%

\author{G\"otz S. Uhrig}%
\email{gu@thp.uni-koeln.de}%
\affiliation{Institut f\"ur Theoretische Physik, Universit\"at zu K\"oln, Z\"ulpicher Str. 77, D-50937 K\"oln, Germany.}%

\begin{abstract}
  We investigate the antiadiabatic limit of an antiferromagnetic
  $\spinonehalf$ Heisenberg chain coupled to Einstein phonons via a bond
  coupling. The flow equation method is used to decouple the spin and the
  phonon part of the Hamiltonian. In the effective spin model longer range
  spin-spin interactions are generated. The effective spin chain is
  frustrated. The resulting temperature dependent couplings are used to
  determine the magnetic susceptibility and to determine the phase transition
  from a gapless state to a dimerized  gapped phase. The susceptibilities and
  the phase diagram obtained via the effective couplings are compared with
  independently calculated quantum Monte Carlo results.
\end{abstract}



\maketitle

\section{Introduction}
\label{secInt}
We study the magnetic susceptibility and the zero-temperature phase diagram
of the antiferromagnetic $\spinonehalf$ Heisenberg chain coupled to
dispersionless Einstein phonons. Due to the coupling of the lattice degrees of
freedom to quasi one-dimensional magnetic degrees of freedom, this model
exhibits a zero-temperature phase transition between a gapless phase and a
massive phase showing dimerization. This quantum phase transition in the
one-dimensional model is regarded as the equivalent to the spin-Peierls
transition\cite{bray83} occurring at finite temperature in the
three-dimensional model. Models of this type were investigated earlier by
Pytte\cite{pytte74b} and Cross and Fisher\cite{cross79}. The discovery of the
first inorganic spin-Peierls substance CuGeO$_3$ \cite{hase93a,bouch96}
renewed the interest in models with spin-phonon (SP) coupling. Various methods
(\eg density matrix renormalization \cite{bursi99}, cutoff renormalization
group \cite{sun00}, linked cluster expansion \cite{trebst99cm}, exact
diagonalization \cite{welle98,weisse99b}, flow equations
\cite{uhrig98b,raas01a}, and quantum Monte Carlo
\cite{sandv97,sandv99,kuehne99}) were used to investigate ground state
properties, especially the phase diagram, the magnetic excitation spectrum and
thermodynamic properties of the spin-phonon chain.

In this paper we compare results of two independent methods applied to the
spin chain Hamiltonian with a local bond coupling to phononic degrees of
freedom: flow equations (continuous unitary transformations) and quantum Monte
Carlo. With the flow equation method we map the initial magneto-elastic
problem onto an effective magnetic problem with temperature dependent
effective spin-spin couplings.
These are used to calculate the magnetic susceptibility (via high temperature
series expansions or exact complete diagonalization) and to determine the
$T=0$ phase separation line. Independently, the phase diagram and $\chi(T)$
are determined by means of quantum Monte Carlo (QMC).

In \tchap \ref{secModelFlow} we introduce the model Hamiltonian for the
spin-phonon chain and present the flow equation approach used. Next, the
quantum Monte Carlo approach is introduced in \tchap \ref{secQMC}. Finally, we
show our results for the $T=0$ phase diagram in \tchap \ref{secPhase} and for
the magnetic susceptibility in \tchap \ref{secSusc}. A short summary is given
in \tchap \ref{secSum}.

\section{Definition of the Model and the Flow Equations}
\label{secModelFlow}
The model under study reads
\begin{subequations}
\label{eqHl}
\begin{align}
  \hl & = \hs + \hsb + \hb\\
  \hs & = \sum_i (J_1 \ldep\spr{}{+1} + J_2\ldep\spr{}{+2})\\
  \hb & = \om \sum_i \bdi\bi \label{eqHsHb}\\
  \hsb & = \sum_i (\opail\bdi+\opadil\bi) \label{eqHlHsb}\ .
\end{align}
\end{subequations}
Here $\ops{}$ stands for the $\spinonehalf$ spin operator on site $i$ and
$\bi$ ($\bdi$) destroys (creates) a phonon on the bond between the sites $i$
and $i+1$. The variable $\ell$ is used to parameterize the flow equation
transformation and is explained in detail in the following paragraph.
$\hs$ is the Hamiltonian of a frustrated spin chain or $J_1$-$J_2$ model. If
the frustration parameter $\al\equiv J_2/J_1$ exceeds a critical value of
$\al=\alc=0.241167(5)$, the model undergoes a quantum phase transition from a
gapless state to a gapped phase \cite{affle89,okamo92,egger96}. We start with
the Hamilton operator (\ref{eqHl}) consisting of pure spin ($\hs$) and phonon
($\hb$) parts and a spin-phonon coupling term $\hsb$. Flow equations are used
to decouple the spin-phonon system.

The flow equation approach was introduced by Wegner in
1994\cite{wegne94,wegne01a}. This method is similar to Fr\"ohlich's approach
\cite{frohl52}, which maps a Hamilton operator onto an effective Hamiltonian in
one step by applying a unitary transformation. Wegner's approach constitutes a
modification of Fr\"ohlich's single step transformation since an infinite
number of infinitesimal unitary transformations is used. The idea behind this
scheme is that the continued adjustment of the infinitesimal transformation to
the Hamiltonian yields a smoother effective interaction than a rotation in one
step.

The continuous unitary transformation is parameterized by a flow parameter
$\ell\in[0,\infty]$. So $\h(\lz)$ stands for the original Hamiltonian.
The effective Hamiltonian $\h(\li)$ is simpler in the way that the direct
spin-phonon coupling has been rotated away. The infinitesimal generator
$\etal$ defines the unitary transformation via
\begin{equation}
  \diffell{\hl}=[\etal,\hl]\ .
  \label{eqFlow}
\end{equation}
A choice of $\etal$ proposed by Wegner is 
\begin{equation}
  \etal=[\hdl,\hl]\ ,
  \label{eqEtaGen}
\end{equation}
where $\hd$ is the part of the Hamiltonian which is taken as the diagonal
part. By choosing $\hdl\equiv\hsl+\hbl$ and the ``off-diagonal'' interaction
part as $\h_{\text{od}}\ldep\equiv\hsbl$, our generator reads
$\etal\equiv[\hd,\h_{\text{od}}] = [\hs+\hb,\hsb]$.
Alternative choices of $\eta$ were used in the literature
\cite{kehre96a,mielk98,knett00a} in order to obtain flow equations with
specific properties. For the present purposes, the canonical choice
(\ref{eqEtaGen}) of $\eta$ is sufficient.

As initial conditions we choose
\begin{equation}
  J_1(0)=J \quad \text{ and } \quad J_2(0)=J_2^0=0.
\end{equation}
So the untransformed pure spin part is the antiferromagnetic $\spinonehalf$
Heisenberg Hamiltonian with a nearest neighbor coupling. In the limit
$\limellinf$ longer range spin-spin interactions arise and the next-nearest
neighbor coupling constant $J_2$ in the effective spin model becomes finite.

$\hsb$ describes the coupling between spin and phonon system. The coupling
operator $\opai(0)$ typically consists of nearest neighbor spin products
since in realistic magnetic materials the spin-phonon coupling mainly
influences neighboring sites. By choosing different coupling operators
various mechanisms of how the lattice distortions influence the exchange
integral $J$ can be investigated. If the exchange coupling depends
on the position of a spin between the neighboring ones the appropriate
choice for $\opai(0)$ is the \emph{difference coupling}
\begin{equation}
  \opai^{\text{diff}}=g(\spr{}{+1}-\spr{}{-1})\ .
  \label{eqDiffCoupling}
\end{equation}
For CuGeO$_3$ this corresponds to changes of two neighboring Cu--O--Cu binding
angles induced by shifts of the copper ion in the middle, \ie one angle is
enlarged at the expense of the other one. This coupling type was investigated
in the framework of a flow equation approach in \trefs
\onlinecite{uhrig98b,raas01a}. In \tref \onlinecite{uhrig98b} an analysis is
presented in leading order in $g\dom$ and in the two leading orders in
$J\dom$. Though only systematic to order $g^2\dom^2$ and $g^2J\dom^3$ a good
qualitative picture for the antiadiabatic limit of the spin-phonon chain could
be achieved. A more subtle analysis is given in \tref
\onlinecite{raas01a}. The price paid in \tref \onlinecite{raas01a} for high
precision results for small $J\dom$ and $g\dom$ is a breakdown of the
formalism for large values of $J\dom$.

In this paper we study the \emph{bond coupling}
\begin{equation}
  \opai(0)=g\spr{}{+1}
  \label{eqOpA}\ ,
\end{equation}
which is frequently used\cite{sandv97,kuehne99,trebst99cm,weisse99b,fehsk00},
especially in connection with CuGeO$_3$. For this type of coupling single
harmonic degrees of freedom directly modify the magnetic interaction. This
is due to side group effects engendered by the germanium atoms
\cite{geert96,khoms96,werne99}. In the present paper we restrict ourselves to
an analysis in the leading orders as in \tref \onlinecite{uhrig98b}.

To study the Hamiltonian (\ref{eqHl}) with the coupling operator (\ref{eqOpA})
via the flow equation method, let us first observe that according to \tref
\onlinecite{wegne94}, $\opai$ should be normal-ordered, \ie
$\opai(0)\rightarrow\opai(0)-\langle\opai(0)\rangle$. In this way the quantum
phonons only couple to the comparably small spin fluctuations and not to the
bare operator $\spr{}{+1}$. This is an extremely important point. Only the
coupling to the small fluctuations justifies the expansion approach. This is
\emph{not} guaranteed by the specific choice (\ref{eqOpA}). This feature of the
coupling type (\ref{eqOpA}) is in contrast to the difference coupling
$\opai^{\text{diff}}$ which does not contribute to the undistorted phase due
to the translational invariance \cite{uhrig98b}, \ie
$\langle\opai^{\text{diff}}(0)\rangle$ vanishes. The necessary normal-ordered
coupling operator for the bond coupling reads
\begin{equation}
  \opai(0)=g(\spr{}{+1}-\NN)
  \label{eqOpAnn}
\end{equation}
where $\NN$ stands for the (temperature dependent) expectation value of the
nearest neighbor spin product $\NNspin$.
The value of $\NN$ is calculated using Pad\'e approximants for the series
expansions given in \trefs \onlinecite{buehler00a,knett00a}.
As we determine $\NN$ in the uniform phase it can be
interpreted as an average value over the chain
For the antiferromagnetic chain the expectation value $\NN$ for
$T=0$ lies between $1/4-\ln(2)\approx-0.443$ (for $\al=0$) and
$3/8=-0.375$ (for the fully dimerized chain).

To achieve normal-ordering we can rewrite the Hamiltonian (\ref{eqHl}) by
shifting the phonons operators via $\bi\to\hat{\bi}-\frac{g\NN}{\om}$. Thus,
the Hamiltonian with the normal-ordered coupling is
\begin{subequations}
  \begin{gather}
    \begin{split}
      \hat{\h} &= \hat{J}\sum_i\spr{}{+1} +
      g\sum_i(\spr{}{+1}-\NN)(\bhdi+\bhi)\\
      &\phantom{= } + \om\sum_i\bhdi\bhi + \sum_i\hat{E}_0
    \end{split}\\
    \hat{J} = J-\frac{2g^2\NN}{\om}
    \quad \text{ and } \quad
    \hat{E}_0 = \frac{g^2\NN^2}{\om}\ .
  \end{gather}
\end{subequations}
We now follow the outline of \tref \onlinecite{uhrig98b}.
Applying the unitary transformation generated by $\eta$, we find an effective
Hamilton operator without spin-phonon coupling terms
\begin{subequations}
\begin{align}
  \begin{split}
    \he &= \hat{J} \sum_i\spr{}{+1} + \Delta\h_{\text{X}} +
    \Delta\h_{\text{Y}}\\
    &\phantom{= } + \om\sum_i\bhdi\bhi + \sum_i\hat{E}_0
  \end{split}\\
  \begin{split}
    \Delta\h_{\text{X}} &= -\frac{1}{\om}\sum_i\opadi\opai\\
    &= \frac{g^2}{2\om}\sum_i \left[ (1+4\NN)\spr{}{+1}
      - 2\NN^2 - 3/8 \right]
  \end{split}\\
  \begin{split}
    \Delta\h_{\text{Y}} &= \frac{1}{2\om^2} \coth\left(\frac{\om}{2T}\right)
    \sum_i\left[\opai,[\hs,\opai]\right]\\
    &= \frac{g^2}{2\om^2}\coth\left(\frac{\om}{2T}\right)\sum_i
    \big[(J_2^0-\hat{J})\spr{}{+1}\\
    &\phantom{= } + (\hat{J}-2J_2^0)\spr{}{+2} + J_2^0\spr{}{+3}\big]\ .
  \end{split}
\end{align}
\end{subequations}
As no frustration is present in the original model (\ie $J_2^0=0$) we
obtain for the effective nearest and next-nearest neighbor couplings
\begin{equation}
  \begin{split}
    J_1=J + \frac{g^2}{2\om} -
    \frac{\hat{J}g^2}{2\om^2}&\coth\left(\frac{\om}{2T}\right)\\
    \quad \text{ and } \quad 
    J_2=\frac{\hat{J}g^2}{2\om^2}&\coth\left(\frac{\om}{2T}\right)
    \label{eqEffCoupl}
  \end{split}
\end{equation}
while longer range two-spin couplings and products with four different spins
are omitted. In this way the dressing of the spins with phonons induces a
frustration $\ale=J_2/J_1>0$. The omission of the longer range and the
four-spin couplings contributes an additional approximation. It is justified by
the fact that these terms contribute neither in a N\'eel state nor in a dimer
state. We note that the couplings are temperature dependent via the
$\coth$ terms and via the $T$-dependence of the expectation value
$\NN=\NNspin$. Though only systematic to order $g^2\dom^2$ and $g^2J\dom^3$,
higher order effects enter due to the appearance of $\hat{J} =
J-\frac{2g^2\NN}{\om}$ in \teqs (\ref{eqEffCoupl}). Since $C$ is calculated
self-consistently, this is in principle a correction of infinite order. In
this way the phonon-shift $\bi\to\hat{\bi}-\frac{g\NN}{\om}$, which is
necessary to apply the flow equation approach, induces infinite order
contributions.

\begin{figure}
\begin{center}
  \includegraphics[width=\linewidth]{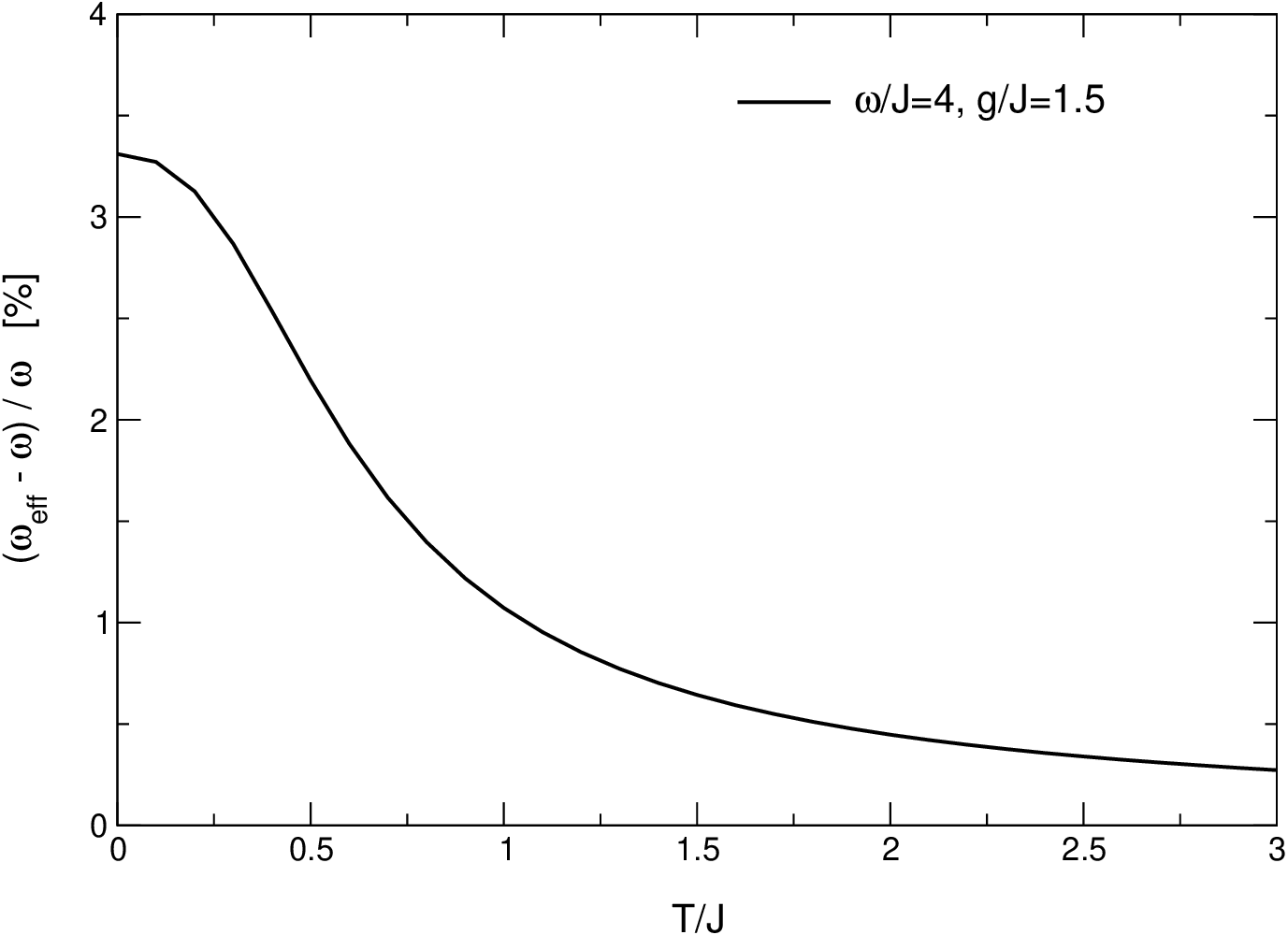}
\end{center}
\caption{\label{figEffOmega}Temperature dependence of the effective phonon
  frequency. The plot shows the relative change in $\om$ for $\om/J=4$
  and $g/J=1.5$.}
\end{figure}

It was pointed out in recent articles \cite{uhrig98b, gros98, sandv99} that the
spin-Peierls transition in the antiadiabatic limit is not accompanied by a
softening of the Peierls-active phonon modes as it is the case for organic
spin-Peierls compounds. This phonon hardening was observed for the difference
coupling using a similar flow equation approach in \tref\onlinecite{raas01a}.
For the bond coupling investigated here a simple formula for the effective
phonon frequency $\om_{\eff}$ can be derived by evaluating the $\bd_{\vec{q}}
b_{\vec{k}}$ term of $\text{d}\Delta\h/\text{d}\ell$ (\cf formula (9) of
\tref\onlinecite{uhrig98b}). Hence we find
\begin{equation}
  \om_{\text{eff}} = \om + \frac{g^2 \hat{J}}{\om^2}
  \left(-\NNspin + \NNNspin\right)
\end{equation}
where non-local terms are neglected. As for zero temperature $\NNspin<0$,
$\NNNspin>0$, and thus $\hat{J} = J-2g^2\NNspin/\om > J$ the effective
phonon frequency is enlarged. With increasing temperature the correlations are
destroyed and accordingly $\om_{\eff}$ tends to $\om$. The generic temperature
dependence of the relative frequency increase is depicted in \tfig
\ref{figEffOmega} for $\om/J=4$ and $g/J=1.5$.

After these remarks regarding the phonon part of the effective Hamiltonian, we
now turn back to the physics in the spin sector. By solving $\al=J_2/J_1=\alc$
for $\gdom$ for $T=0$ we obtain an explicit formula for the phase separation
line
\begin{equation}
    \frac{\gc^2}{\om^2} = \frac{\gac - \sqrt{(\gac-\alc)^2 - 16\NN\alc\gac} -
    \alc}{4\NN(1+\alc)}
  \label{eqAlphaC}
\end{equation}
with the abbreviation $\gac=(1+\alc)\jdom$. The two limiting cases are then
given by
\begin{subequations}
\begin{align}
  \lim_{\jdom\to0}\frac{\gc}{\om} &=
  \sqrt{\frac{-\alc}{2\NN(1+\alc)}}\approx 0.4682\\
  \lim_{\jdom\to\infty}\frac{\gc}{\om} &=
  \sqrt{\frac{2\alc}{1+\alc}}\approx 0.6234\ .
\end{align}
\end{subequations}
The finiteness of the critical spin-phonon coupling $\gcdom$ for $\jdom\to0$
is a striking difference between the local coupling mechanism considered here
and the coupling to the order parameter used \eg in \tref
\onlinecite{raas01a}. A qualitative understanding can be achieved by
considering the shifted and rescaled Hamiltonian
\begin{align}
\begin{split}
  \frac{\h}{\om} = &
  \left[ \frac{J}{\om}
    - 2 \left(\frac{g}{\om}\right)^2\NN \right] \sum_i \spr{}{+1}
  + \sum_i \bhdi\bhi\\
  & + \sum_i\frac{\hat{E}_0}{\om}
  + \frac{g}{\om} \sum_i \left(\spr{}{+1}-\NN\right)
  \left(\bhdi+\bhi\right)\ .
\end{split}
\end{align}
If $g\dom$ is finite the effective nearest neighbor coupling constant remains
finite even in the limit $J\dom\to0$, as the expectation value $C$ is
typically $\approx-0.4$. Thus, there is no reason why the critical spin-phonon
coupling should vanish for $J\dom$ approaching zero. This illustrates an
important difference between the difference coupling (\ref{eqDiffCoupling})
and the bond coupling (\ref{eqOpA}).

\begin{figure}
\begin{center}
  \includegraphics[width=\linewidth]{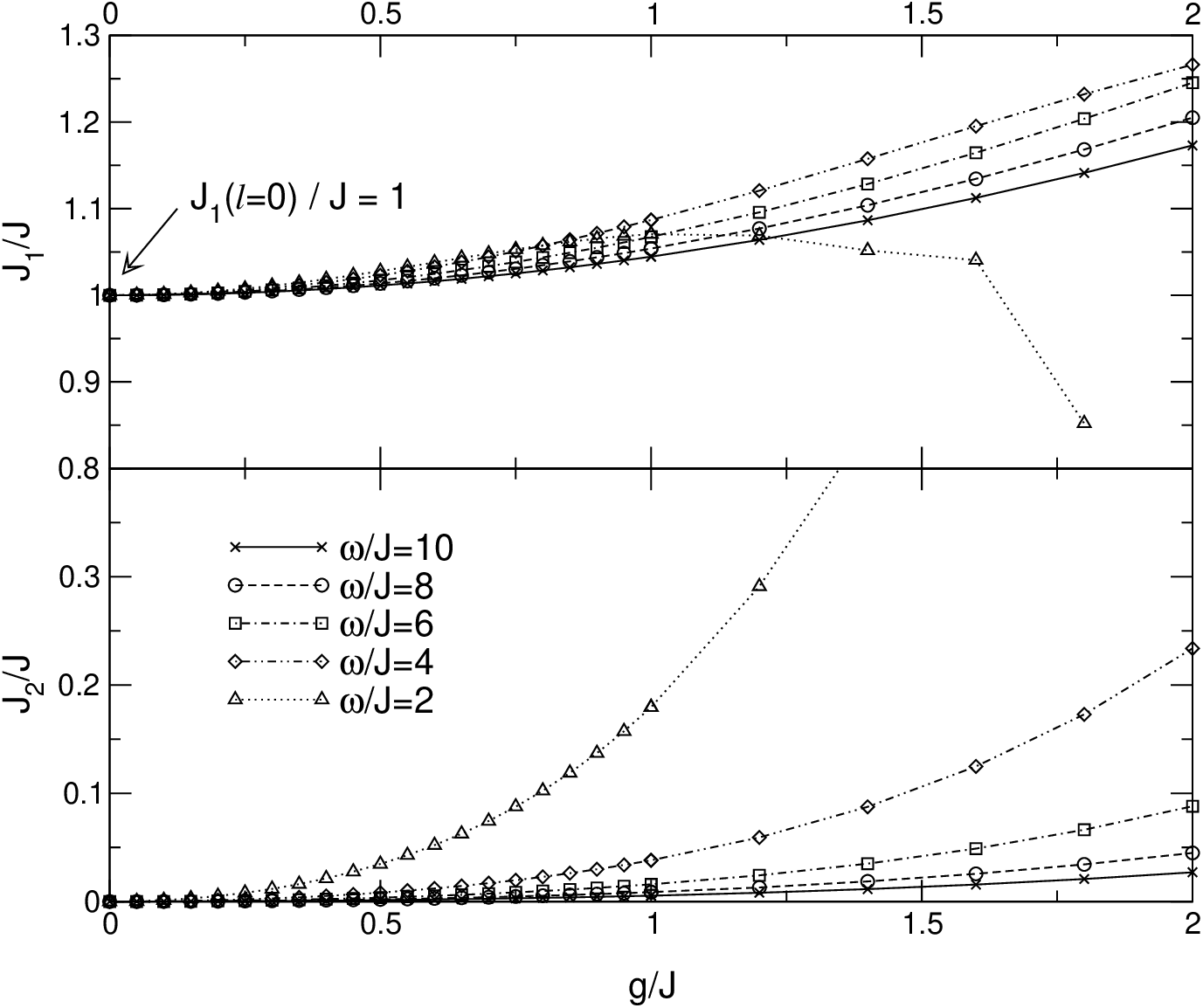}
\end{center}
\caption{\label{figEffCouplings}Dependence of the effective nearest and
  next-nearest neighbor spin-spin couplings from the spin-phonon coupling $g$
  for fixed values of $\om$ and zero temperature.}
\end{figure}
\tfig \ref{figEffCouplings} shows the change of the effective
spin-spin couplings with increasing spin-phonon coupling. Since our
method corresponds to the leading orders of an expansion in $J\dom$ and
$g\dom$, the approach breaks down if the ratio $J\dom$ becomes too large. This
can be seen in the unsystematic downturn of the $J_1(g)$ curve for $\om=2J$.

To obtain the magnetic susceptibilities for the spin-phonon chain we follow
various routes. Using the effective couplings (\ref{eqEffCoupl}) we perform an
exact complete diagonalization for the frustrated spin chain or use
alternatively high temperature expansions \cite{buehler00a}. In order to
extrapolate the results we use Dlog-Pad\'e approximants
\cite{uhrig00a,buehler01a}. If the spin-phonon coupling $g\dom$ is small, the
effective frustration is negligible. This justifies the third approach to
simply rescale the exact results for the \emph{unfrustrated} Heisenberg chain
\cite{klump93b,egger94,klump98a} by the renormalized nearest-neighbor coupling
$J_1$. The results for these three methods agree perfectly well with the
direct QMC data, which are \emph{not} based on the derived effective
couplings.

\section{Quantum Monte Carlo}
\label{secQMC}
In a completely independent approach we study the model (\ref{eqHl}) directly
by a quantum Monte Carlo method based on the loop algorithm \cite{evertz93}.
The technical details of the method are described in \tref
\onlinecite{kuehne99}. Here we just point out again, that the algorithm allows
to incorporate as many phonons as desired. We usually choose 50 per site. Thus
our results are not hampered by too low cutoffs in the phonon number, which
represent otherwise a serious difficulty at low phonon frequencies.  As has
been demonstrated in \tref \onlinecite{kuehne99} the error originating from
the finiteness of the Trotter number can be easily overcome by extrapolation.
Here Trotter numbers of $M=40$, $60$ and $80$ are used.  The calculations are
performed with $10^5$--$10^6$ spin updates and $20$ phonon updates per spin
update. We considered up to $256$ sites so that the finite-size effects are
controllably small. Thus, apart from statistical fluctuations the Monte Carlo
results are exact.

It was appropriate for the quantum Monte Carlo study to use the Hamiltonian in
the following form \cite{kuehne99}
\begin{equation}
  \tilde{\h} =
  2\tilde{J}\sum_i(\spr{}{+1}-\frac{1}{4})[1+\tilde{g}(\bdi+\bi)]
  + \tilde{\om}\sum_i\bdi\bi
  \label{eqHqmc}
\end{equation}
where we use the tilde to distinguish clearly between the different coupling
constants in the different ways to denote the Hamiltonian. The operator
(\ref{eqHqmc}) can be mapped onto (\ref{eqHl}) by a shift in the phonon
operators
\begin{equation}
  \bi\to\bi+\frac{\tilde{J}\tilde{g}}{2\tilde{\om}}\ .
  \label{eqShift}
\end{equation}
By this mapping one obtains
\begin{align}
  \begin{split}
    \label{eqHqmcshifted}
    \tilde{\h} = &
    2\tilde{J}(1+\frac{\tilde{J}\tilde{g}^2}{\tilde{\om}}) \sum_i\spr{}{+1} +
    \tilde{\om}\sum_i\bdi\bi +\\
    & 2\tilde{g}\tilde{J} \sum_i\spr{}{+1}(\bdi+\bi) -
    \frac{\tilde{J}}{2}(1-\frac{\tilde{J}\tilde{g}^2}{2\tilde{\om}})\ .
  \end{split}
\end{align}
Accounting for this rescaling of the spin-spin coupling constant the
susceptibilities computed for $\tilde{\h}$ in \teq (\ref{eqHqmc}) and those
for $\h$ in \teq (\ref{eqHl}) can be compared.

The quantum Monte Carlo procedure is per construction a method giving results
at finite temperature $T$. But since the algorithm allows to approach very
low $T$ one can also study properties of the ground state, in particular if
one considers the model in its gapped phase. This was used in the following
to extract the phase boundaries of the model.

\section{Results}

\subsection{Phase diagram}
\label{secPhase}
\begin{figure}
\begin{center}
  \includegraphics[width=\linewidth]{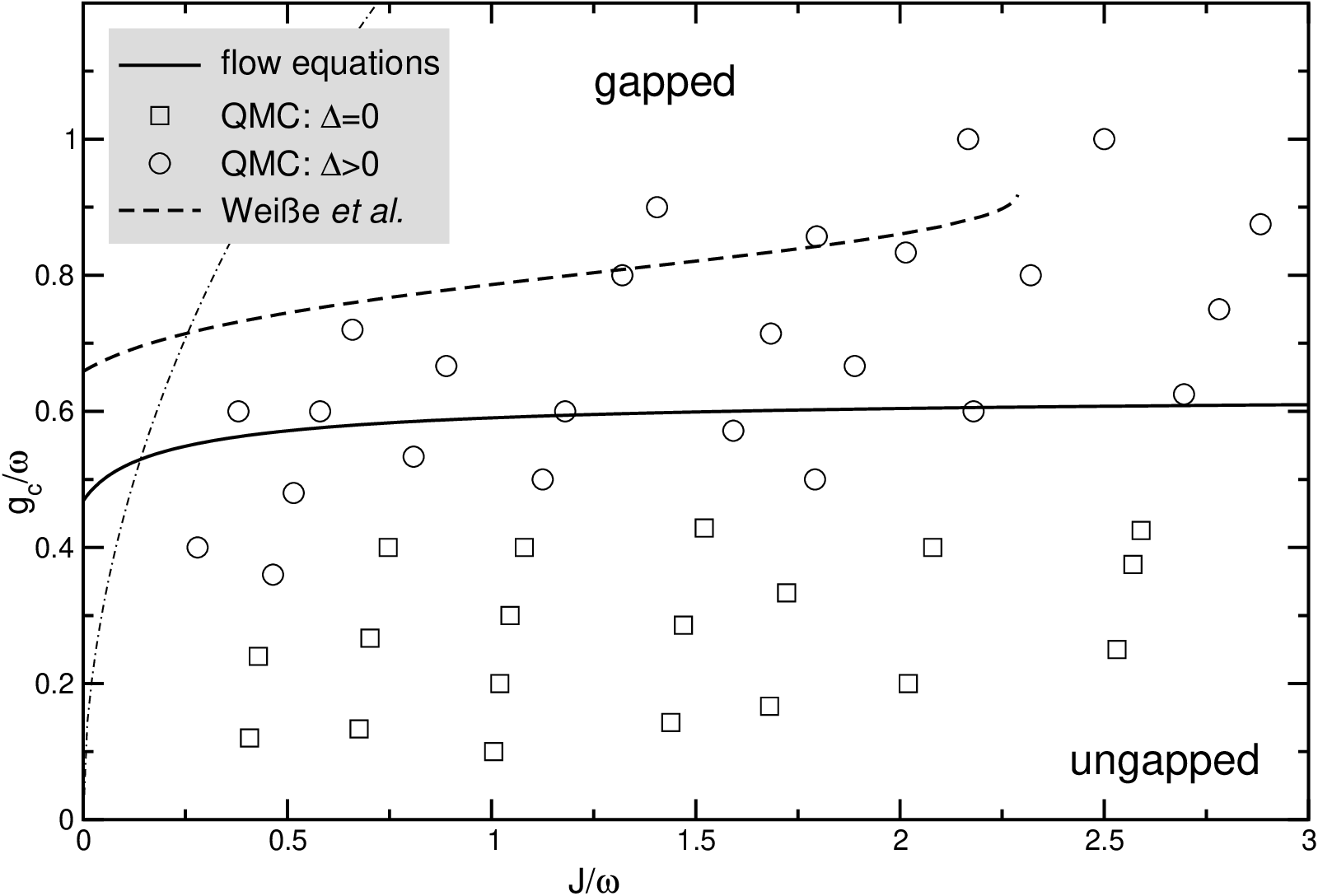}
\end{center}
\caption{\label{figPHASE}Zero temperature phase diagram of the spin-Peierls
  antiferromagnetic chain of spins interacting with phonons. For small values
  of the spin-phonon coupling $g\dom$ the system is gapless ($\Delta=0$). For
  large $g\dom$ the system is dimerized and has an energy gap ($\Delta>0$).
  The circles (squares) correspond to QMC data, where (no) dimerization was
  found for $T/J=0.05$ and a chain of $N=256$ sites. QMC data can only be
  computed to the right of the dot-dashed line (for details see main text).
  The results from Wei\ss{}e \tetal are taken from \tref
  \onlinecite{weisse99b}.}
\end{figure}
The calculation via flow equations is based on the idea that the transition
into the ordered phase does not occur due to the softening of a phonon but due
to the tendency of the effective spin model towards dimerization.
Phonon-induced frustration above its critical value $\ale>\alc$ drives the
dimerization. The phase transition line is therefore determined by solving
\begin{eqnarray}
  \alc = \big(\lim_{\limellinf}\al(g)\big)\big|_J
\end{eqnarray}
for fixed values of $J$ with respect to $g$ (\cf \teq \ref{eqAlphaC}). In this
way we find the critical spin-phonon coupling $\gc$ in dependence of the
nearest neighbor spin coupling $J$ as depicted in \tfig \ref{figPHASE}.

\begin{figure}
\begin{center}
  \includegraphics[width=\linewidth]{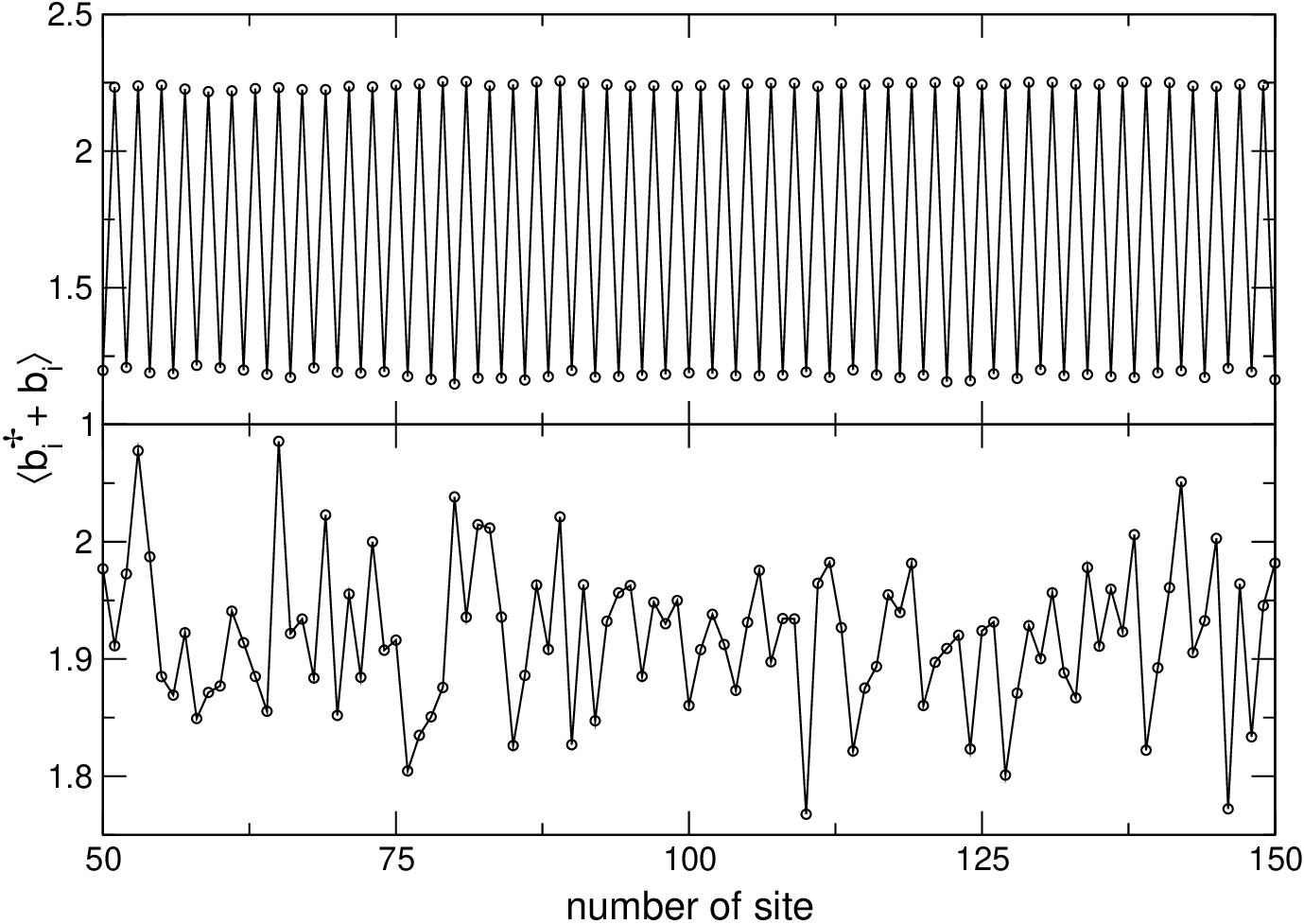}
\end{center}
\caption{\label{figDisplacement}Phonon displacement averaged over $10^4$
  configurations for $\tilde{\om}=10J$ and $\tilde{g}=0.8$
  ($\tilde{g}=0.1$) for a system size of $N=256$ sites in the upper (lower)
  plot. For $\tilde{g}=0.8$ the dimerization is clearly visible.}
\end{figure}

Using the quantum Monte Carlo method the decision whether a pair
$(J\dom,g\dom)$ represents a point in the ungapped or the gapped phase is
reached by investigating the expectation values of the local
phonon displacements $\langle\bdi+\bi\rangle$. At low temperatures $\langle
\bdi+\bi\rangle$ show clear dimerization patterns if the model is in the
gapped phase and they are randomly fluctuating in the critical phase of the
model (see \tfig \ref{figDisplacement}). Of course, the ergodicity of the Monte
Carlo procedure complicates the identification of the ground state in the
gapped phase since for long run times the ground state fluctuates always
between the two possible dimerized patterns. However, by averaging over not too
many configurations (we generally take 10000) one can extract well the
dimerization patterns from the Monte Carlo results.

For intermediate values of $J\dom$ the flow equation and QMC method predict
similar values for the strength of the critical coupling $\gc\dom$ (\cf \tfig
\ref{figPHASE}). There is a certain tendency that the QMC data indicates lower
values for $\gc\dom$ than the analytical calculation. We attribute this
discrepancy partly to the approximate character of the analytical calculation
and partly to the difficulty to determine the phase boundary by QMC precisely.
Note that due to the phonon shift (\ref{eqShift}) the QMC approach is
restricted to the region on the right hand side of the dot-dashed line in
\tfig \ref{figPHASE}. The QMC data indicate a smaller value for $\gc\dom$ in
the antiadiabatic limit. In the immediate vicinity of the dot-dashed line (\ie
for very large $\tilde{\om}$) it is rather hard to distinguish between a fully
dimerized pattern and partially dimerized structures. Thus, further
investigations are desirable for the extreme antiadiabatic limit of the phase
diagram.

The phase separation line predicted by Wei\ss{}e \tetal \cite{weisse99b} via
fourth order perturbation theory lies well above the flow equation curve and
above the Monte Carlo results. This was also observed in \tref
\onlinecite{raas01a} where in the case of the difference coupling the leading
order result\cite{uhrig98b} compared much better with the DMRG
data\cite{bursi99} than the fourth order result \cite{weisse99b}. Since with a
difference coupling the critical spin-phonon coupling $\gc\dom$ vanishes as
$J\dom$ goes to zero the expansion in these parameters becomes exact. Thus,
the phase separation lines for the difference coupling coincide for small
values of $J\dom$.\cite{raas01a} But as discussed above, in the case of a bond
coupling $\gc\dom$ is \emph{non-zero} in the antiadiabatic limit
($J\dom\to0$). Hence, it is not surprising that even in the limit of vanishing
$J\dom$ the deviation does not decrease. \tref \onlinecite{weisse99b} predicts
even a value of $\gc\dom=\sqrt{8/3\cdot\alc/(1+2\alc)}\approx0.6587$ for
$J\dom\to0$.

In \tref \onlinecite{sandv99} a QMC analysis of the spin-phonon chain with a
bond coupling was presented for the parameter $J\dom=4$. A critical coupling
$\gc\dom=0.636\pm0.042$ was calculated by investigating the staggered spin
susceptibility for various system sizes. This result is in very good agreement
with $\gc\dom=0.613$ predicted by the flow equation method (not depicted in the
phase diagram \tfig \ref{figPHASE}).

\subsection{Magnetic susceptibility}
\label{secSusc}
In this \tchapter we consider the influence of the spin-phonon coupling on a
thermodynamic quantity like the magnetic susceptibility $\chi(T)$. We show
that the magnetic susceptibility can be well described by an effective spin
model. Furthermore, the question whether it is possible to describe
thermodynamic quantities of the SP chains with a static spin model or via a
spin model with temperature dependent couplings $J(T)$ is investigated. The
exchange integral $J$ of a quantum spin system is often assumed to be
constant. However, in real materials where $J$ depends on the actual positions
of the magnetic ions in the crystal this approximation need not be valid. For
the spin-Peierls substance CuGeO$_3$ this point is of special interest. The
magnetic susceptibility of CuGeO$_3$ in the temperature range above the
spin-Peierls transition can be well fitted by a frustrated Heisenberg
model.\cite{casti95,riera95,fabri98a} It is often objected, however, that the
agreement might be accidental as no interchain interactions and spin-phonon
couplings are taken into account. This motivates us --- although the
Hamiltonian introduced in \teq (\ref{eqHl}) is too simple to be a realistic
model for CuGeO$_3$ --- to study in detail the influence of the spin-phonon
coupling on the thermodynamic observables. Another substance with strong
magneto-elastic couplings, (VO)$_2$P$_2$O$_7$, was investigated recently
\cite{uhrig00a} by a flow equation approach. The authors found that even for
strong spin-phonon coupling the influence of phonons on thermodynamic magnetic
quantities is rather small. A good description of such quantities could
already be achieved by a static model.

The flow equation approach maps the spin-phonon chain onto a frustrated spin
chain with temperature dependent spin-spin couplings. The corresponding
$\chi(T)$ is compared with the susceptibility calculated with QMC for the full
Hamiltonian (\ref{eqHqmc}). This enables us to study the influence of the
spin-phonon coupling on $\chi(T)$. Furthermore, we can discuss the question
whether it is possible to retrieve the main physical properties of a
spin-phonon chain by looking at its effective spin model.

\begin{figure}
\begin{center}
  \includegraphics[width=\linewidth]{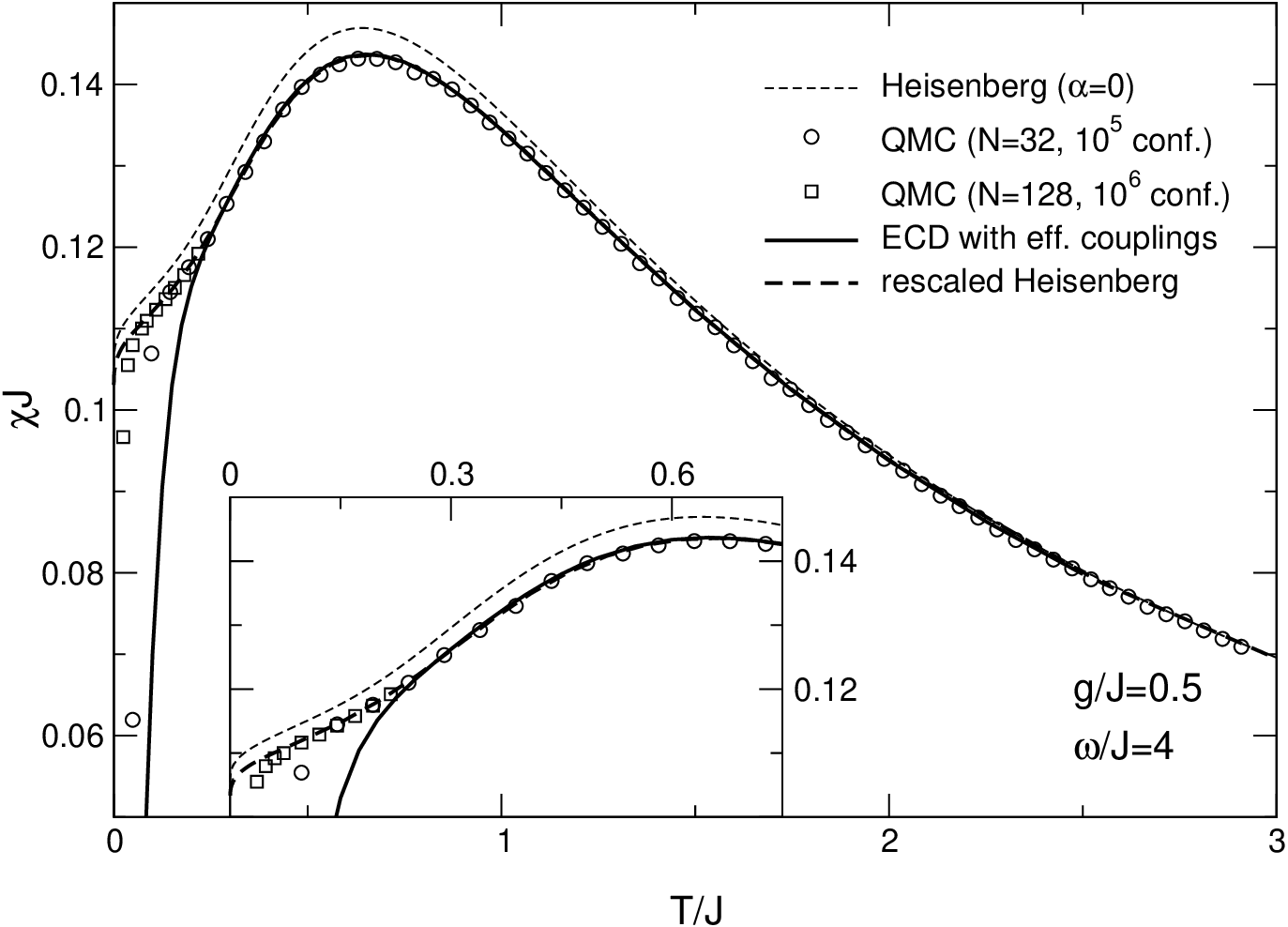}
\end{center}
\caption{\label{figSUSg05om04}Magnetic susceptibility for $g=0.5J$ and $\om=4J$.}
\end{figure}
\begin{figure}
\begin{center}
  \includegraphics[width=\linewidth]{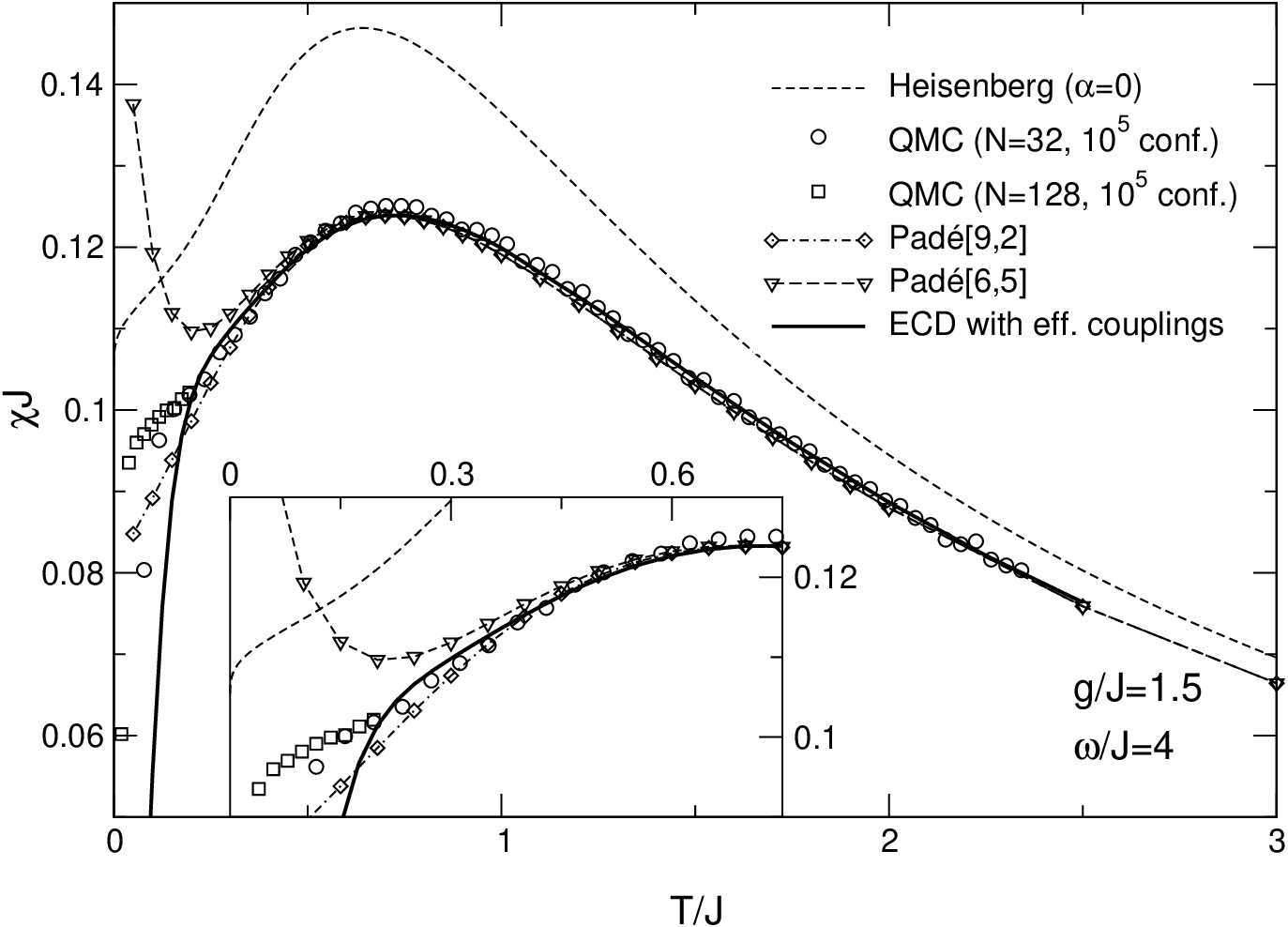}
\end{center}
\caption{\label{figSUSg15om04}Magnetic susceptibility for $g=1.5J$ and $\om=4J$.}
\end{figure}
\begin{figure}
\begin{center}
  \includegraphics[width=\linewidth]{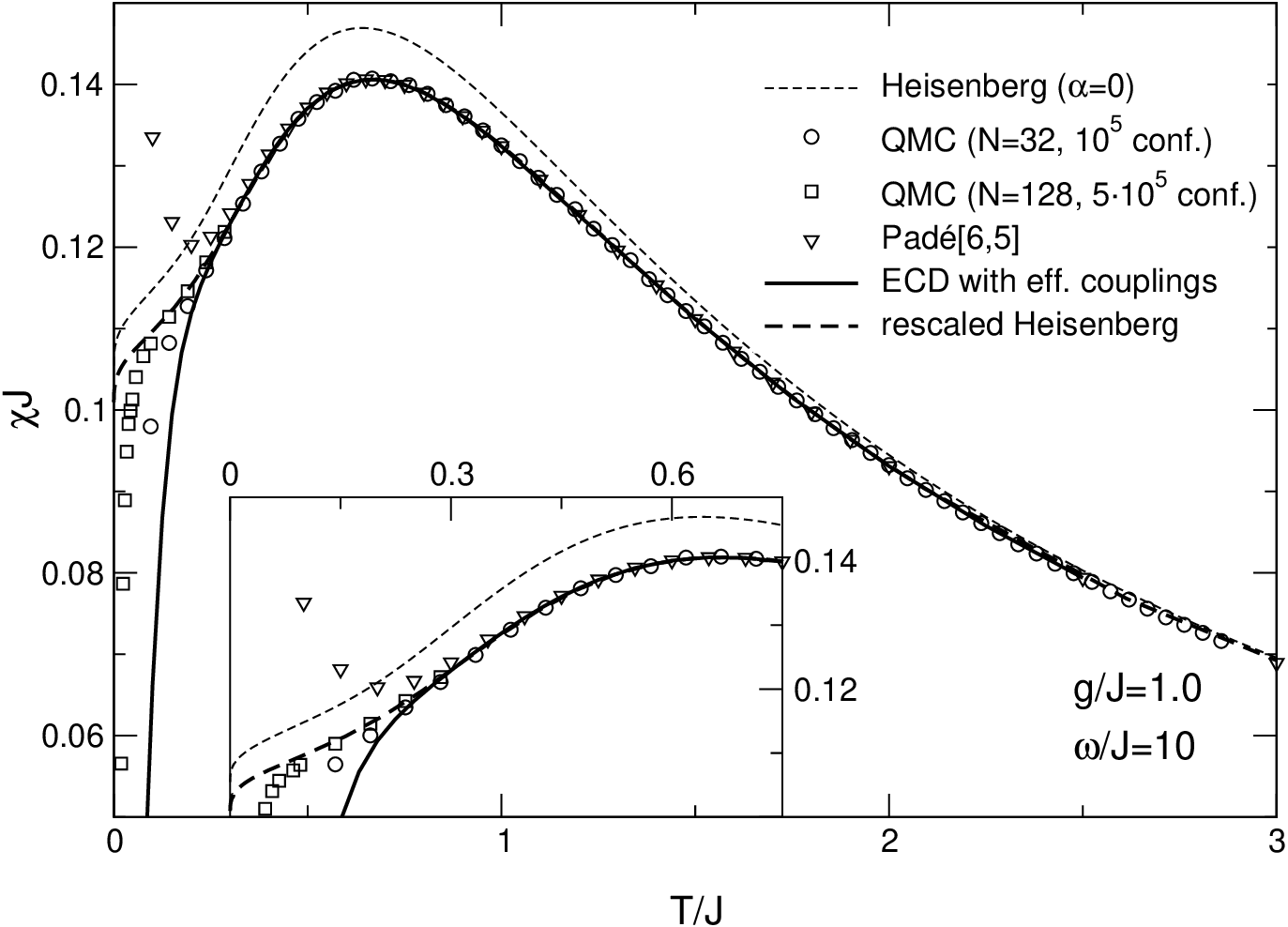}
\end{center}
\caption{\label{figSUSg10om10}Magnetic susceptibility for $g=1.0J$ and $\om=10J$.}
\end{figure}
\begin{figure}
\begin{center}
  \includegraphics[width=\linewidth]{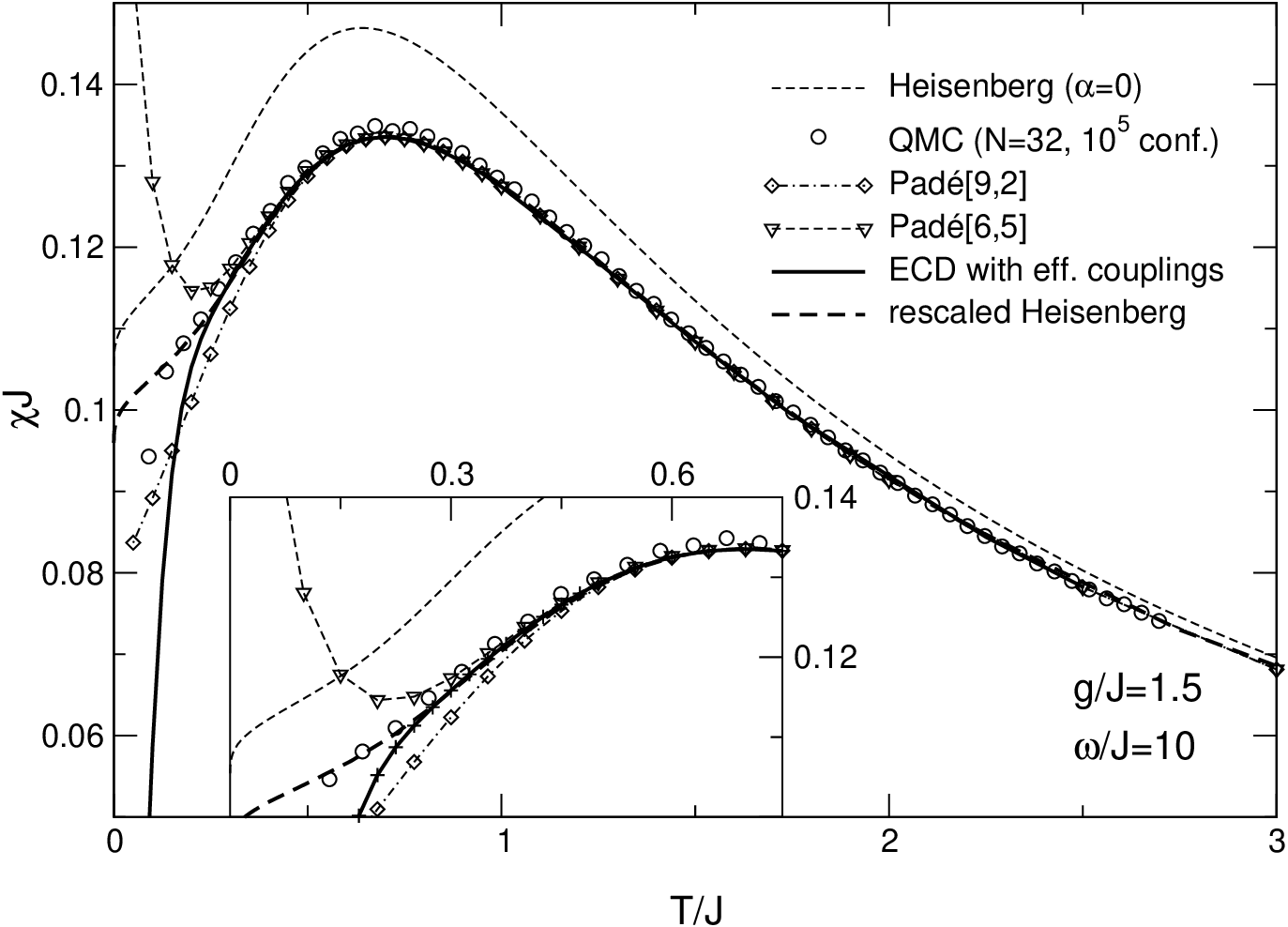}
\end{center}
\caption{\label{figSUSg15om10}Magnetic susceptibility for $g=1.5J$ and $\om=10J$.}
\end{figure}

Figures \ref{figSUSg05om04}-\ref{figSUSg15om10} show our results for four
different values of $g$ and $\om$. All parameters are given according to the
Hamiltonian (\ref{eqHl}). As a reference the exact result of the
unfrustrated Heisenberg chain \cite{klump93b,egger94,klump98a} is depicted in
all figures. \tfig \ref{figSUSg05om04} illustrates that a rather small
spin-phonon coupling $g=0.5J$ leads only to a lowering of the overall height
of the susceptibility. The frustration in the effective model is very small
($J_2/J_1\approx0.008$ for $T=0$ and $J_2/J_1\approx0.02$ for $T/J=5$). This
enables us to rescale the Heisenberg curve with respect to the enlarged
coupling constant $J_1$. To ensure that the effective frustration is really
small enough to be neglected we make an additional check via the exact and
complete diagonalization (ECD) data (computed with the temperature dependent
couplings $J_1$ and $J_2$) for a frustrated chain with $N=16$ sites. The QMC
and the flow equation results agree perfectly down to very low temperatures.

When $g$ is increased the maximum decreases as can be seen in \tfig
\ref{figSUSg15om04} for $\om=4J$ and $g=1.5J$.
Again, the two approaches yield the same results down to $T/J\approx0.2$. The
frustration is no longer negligible. As one can see, the use of an effective
spin model to describe $\chi(T)$ for a spin chain coupled to phonons is well
justified. An easy way to fit experimental data is to use Pad\'e approximants
for high temperature series expansion instead of exact complete
diagonalization. The maximum is well-described by this procedure, even though
the limit of very low temperatures is problematic.

Figures \ref{figSUSg10om10} and \ref{figSUSg15om10} show additional results
for $g=1.0J$ and $g=1.5J$ and for $\om=10J$. As $J\dom$ is small the flow
equation approach works especially well and thus the agreement is even better
compared with the $\om=4J$ results.

Three final remarks regarding the susceptibility results are in order.
In \tref \onlinecite{kuehne99}, the authors observed that with increasing
spin-phonon coupling the susceptibility curves are shifted to higher
temperatures. The main reason for this is the static term $\propto
\langle\bdi+\bi\rangle$ in the original QMC Hamiltonian (\ref{eqHqmc}). This
is not visible in our way of depicting $\chi(T)$ using the parameterization
(\ref{eqHl}) as we shifted the phonons to make this term vanish. If this
static effect is accounted for by the phonon shift, the positions of the
maxima of $\chi(T)$ are nearly the same for all parameters $g$ and $\om$
presented here. Only if $g\dom$ and thus the effective frustration are very
large a small shift due to the temperature dependent frustration becomes
visible.

The second remark concerns the applicability of the flow equation procedure
presented here. As we have already seen in \tfig \ref{figEffCouplings}, the
ratio $J\dom$ must not become too large. Thus, we are only able to show
susceptibilities with $J\dom\lesssim1/4$. The main problem for a thermodynamic
quantity is the $\coth(\om/2T)$ term in the formula (\ref{eqEffCoupl}) for the
effective couplings because it diverges for $T\to\infty$. To reach higher
values of $J\dom$ a more subtle flow equation approach would be necessary. But
even the straightforward flow equation results agree well with the QMC data
for not too large $J\dom$ and $g\dom$.

The third remark concerns the temperature dependence of the effective
couplings. This dependence is fairly small for the parameter values in \tfigs
\ref{figSUSg05om04}-\ref{figSUSg15om10} which is due to
the small values $J\dom$. Thus good fits are also obtained neglecting the
temperature dependence $J(T)\equiv J$ in a model of \emph{static} effective
couplings. This is in agreement with previous results \cite{uhrig00a}. But we
expect that for larger values of $J\dom$ the temperature dependence will be
important. Yet such parameters are beyond the scope of the present approach
based on the leading orders. Hence further investigations of this issues are
called for.

\section{Summary}
\label{secSum}
We applied two methods to study the phase diagram and the magnetic
susceptibility of the isotropic antiferromagnetic spin $1/2$ Heisenberg model
coupled to Einstein phonons via a bond coupling: the flow equation method and
a quantum Monte Carlo approach.

The flow equation approach maps the full problem onto an effective magnetic
problem, \ie onto a frustrated spin chain. The calculation was performed in
leading order in $g\dom$ and in the two leading orders in $J\dom$ and
is thus systematic to order $g^2\dom^2$ and $g^2J\dom^3$. The expansion
approach works especially well in the antiadiabatic limit.

To apply the flow equation formalism, it is important to use the spin operator
which couples to the phonons in its normal-ordered form. To obtain $\chi(T)$
from the calculated effective couplings $J_1(T)$ and $J_2(T)$ we used high
temperature series expansions and exact complete diagonalization.

The quantum Monte Carlo calculations are completely independent of the
effective couplings and are based on a loop algorithm. Spin and phonon degrees
of freedom are dealt with an equal footing.

The zero temperature phase diagram derived by these methods agrees well for
intermediate $J\dom$ and is in agreement with the result in \tref
\onlinecite{sandv99}. Approaching the limit $J\dom\to0$ we find a finite
critical spin-phonon coupling $\gc\dom$ as an important difference to the
difference coupling. This is rigorous since a vanishing $\gc\dom$ for $J\dom$
would correspond to the limit where our approach becomes exact. We cannot fix
the exact value of $\gc\dom$ in this limit as the flow equation approach
corresponds to a leading order expansion in $J\dom$ and $g\dom$ and thus is
not exact for finite $g\dom$. Yet both our approaches yield a significantly
lower value of $\gc\dom$ than the fourth order calculation \cite{weisse99b}.

We compared the magnetic susceptibilities obtained by both methods and find
good agreement. We conclude that at least in this parameter regime the
thermodynamic properties of the Hamiltonian \teq \ref{eqHl} can be described
excellently by an effective spin model with temperature dependent couplings.
The restriction in $g\dom$ and $J\dom$ for the the flow equations are due the
leading order character of the approach presented here and could be overcome
by a more advanced analysis. The flow equation method provides a good
completion to an elaborate QMC analysis as it can be used more quickly, so
that the fitting of parameters is easier.

\begin{acknowledgments}
  The authors acknowledge helpful discussions with A.\@ B\"uhler, K.\@
  Fabricius, H.\@ Fehske, A.\@ Kl\"umper, E.\@ M\"uller-Hartmann, and
  A.\@ Wei\ss{}e. RK and CR  were supported by the Deutsche
  Forschungsgemeinschaft (DFG). GSU acknowledges funding by the DFG in the
  Schwerpunkt 1073.
\end{acknowledgments}


\begin{thebibliography}{39}
\expandafter\ifx\csname natexlab\endcsname\relax\def\natexlab#1{#1}\fi
\expandafter\ifx\csname bibnamefont\endcsname\relax
  \def\bibnamefont#1{#1}\fi
\expandafter\ifx\csname bibfnamefont\endcsname\relax
  \def\bibfnamefont#1{#1}\fi
\expandafter\ifx\csname citenamefont\endcsname\relax
  \def\citenamefont#1{#1}\fi
\expandafter\ifx\csname url\endcsname\relax
  \def\url#1{\texttt{#1}}\fi
\expandafter\ifx\csname urlprefix\endcsname\relax\def\urlprefix{URL }\fi
\providecommand{\bibinfo}[2]{#2}
\providecommand{\eprint}[2][]{\url{#2}}

\bibitem[{\citenamefont{Bray et~al.}(1983)\citenamefont{Bray, Interrante,
  Jacobs, and Bonner}}]{bray83}
\bibinfo{author}{\bibfnamefont{J.~W.} \bibnamefont{Bray}},
  \bibinfo{author}{\bibfnamefont{L.~V.} \bibnamefont{Interrante}},
  \bibinfo{author}{\bibfnamefont{I.~S.} \bibnamefont{Jacobs}},
  \bibnamefont{and} \bibinfo{author}{\bibfnamefont{J.~C.}
  \bibnamefont{Bonner}}, in \emph{\bibinfo{booktitle}{Extended Linear Chain
  Compounds}}, edited by \bibinfo{editor}{\bibfnamefont{J.~S.}
  \bibnamefont{Miller}} (\bibinfo{publisher}{Plenum Press},
  \bibinfo{address}{New York}, \bibinfo{year}{1983}), vol.~\bibinfo{volume}{3},
  p. \bibinfo{pages}{353}.

\bibitem[{\citenamefont{Pytte}(1974)}]{pytte74b}
\bibinfo{author}{\bibfnamefont{E.}~\bibnamefont{Pytte}},
  \bibinfo{journal}{Phys. Rev. B} \textbf{\bibinfo{volume}{10}},
  \bibinfo{pages}{4637} (\bibinfo{year}{1974}).

\bibitem[{\citenamefont{Cross and Fisher}(1979)}]{cross79}
\bibinfo{author}{\bibfnamefont{M.~C.} \bibnamefont{Cross}} \bibnamefont{and}
  \bibinfo{author}{\bibfnamefont{D.~S.} \bibnamefont{Fisher}},
  \bibinfo{journal}{Phys. Rev. B} \textbf{\bibinfo{volume}{19}},
  \bibinfo{pages}{402} (\bibinfo{year}{1979}).

\bibitem[{\citenamefont{Hase et~al.}(1993)\citenamefont{Hase, Terasaki, and
  Uchinokura}}]{hase93a}
\bibinfo{author}{\bibfnamefont{M.}~\bibnamefont{Hase}},
  \bibinfo{author}{\bibfnamefont{I.}~\bibnamefont{Terasaki}}, \bibnamefont{and}
  \bibinfo{author}{\bibfnamefont{K.}~\bibnamefont{Uchinokura}},
  \bibinfo{journal}{Phys. Rev. Lett.} \textbf{\bibinfo{volume}{70}},
  \bibinfo{pages}{3651} (\bibinfo{year}{1993}).

\bibitem[{\citenamefont{Boucher and Regnault}(1996)}]{bouch96}
\bibinfo{author}{\bibfnamefont{J.~P.} \bibnamefont{Boucher}} \bibnamefont{and}
  \bibinfo{author}{\bibfnamefont{L.~P.} \bibnamefont{Regnault}},
  \bibinfo{journal}{J. Phys. I France} \textbf{\bibinfo{volume}{6}},
  \bibinfo{pages}{1939} (\bibinfo{year}{1996}).

\bibitem[{\citenamefont{Bursill et~al.}(1999)\citenamefont{Bursill, McKenzie,
  and Hamer}}]{bursi99}
\bibinfo{author}{\bibfnamefont{R.~J.} \bibnamefont{Bursill}},
  \bibinfo{author}{\bibfnamefont{R.~H.} \bibnamefont{McKenzie}},
  \bibnamefont{and} \bibinfo{author}{\bibfnamefont{C.~J.} \bibnamefont{Hamer}},
  \bibinfo{journal}{Phys. Rev. Lett.} \textbf{\bibinfo{volume}{83}},
  \bibinfo{pages}{408} (\bibinfo{year}{1999}).

\bibitem[{\citenamefont{Sun et~al.}(2000)\citenamefont{Sun, Schmeltzer, and
  Bishop}}]{sun00}
\bibinfo{author}{\bibfnamefont{P.}~\bibnamefont{Sun}},
  \bibinfo{author}{\bibfnamefont{D.}~\bibnamefont{Schmeltzer}},
  \bibnamefont{and} \bibinfo{author}{\bibfnamefont{A.~R.}
  \bibnamefont{Bishop}}, \bibinfo{journal}{Phys. Rev. B}
  \textbf{\bibinfo{volume}{62}}, \bibinfo{pages}{11308} (\bibinfo{year}{2000}).

\bibitem[{\citenamefont{Trebst et~al.}()\citenamefont{Trebst, Elstner, and
  Monien}}]{trebst99cm}
\bibinfo{author}{\bibfnamefont{S.}~\bibnamefont{Trebst}},
  \bibinfo{author}{\bibfnamefont{N.}~\bibnamefont{Elstner}}, \bibnamefont{and}
  \bibinfo{author}{\bibfnamefont{H.}~\bibnamefont{Monien}},
  \bibinfo{howpublished}{\textbf{cond-mat}/9907266 (1999)}.

\bibitem[{\citenamefont{Wellein et~al.}(1998)\citenamefont{Wellein, Fehske, and
  Kampf}}]{welle98}
\bibinfo{author}{\bibfnamefont{G.}~\bibnamefont{Wellein}},
  \bibinfo{author}{\bibfnamefont{H.}~\bibnamefont{Fehske}}, \bibnamefont{and}
  \bibinfo{author}{\bibfnamefont{A.~P.} \bibnamefont{Kampf}},
  \bibinfo{journal}{Phys. Rev. Lett.} \textbf{\bibinfo{volume}{81}},
  \bibinfo{pages}{3956} (\bibinfo{year}{1998}).

\bibitem[{\citenamefont{Wei\ss{}e et~al.}(1999)\citenamefont{Wei\ss{}e,
  Wellein, and Fehske}}]{weisse99b}
\bibinfo{author}{\bibfnamefont{A.}~\bibnamefont{Wei\ss{}e}},
  \bibinfo{author}{\bibfnamefont{G.}~\bibnamefont{Wellein}}, \bibnamefont{and}
  \bibinfo{author}{\bibfnamefont{H.}~\bibnamefont{Fehske}},
  \bibinfo{journal}{Phys. Rev. B} \textbf{\bibinfo{volume}{60}},
  \bibinfo{pages}{6566} (\bibinfo{year}{1999}).

\bibitem[{\citenamefont{Uhrig}(1998)}]{uhrig98b}
\bibinfo{author}{\bibfnamefont{G.~S.} \bibnamefont{Uhrig}},
  \bibinfo{journal}{Phys. Rev. B} \textbf{\bibinfo{volume}{57}},
  \bibinfo{pages}{14004} (\bibinfo{year}{1998}).

\bibitem[{\citenamefont{Raas et~al.}(2001)\citenamefont{Raas, B\"uhler, and
  Uhrig}}]{raas01a}
\bibinfo{author}{\bibfnamefont{C.}~\bibnamefont{Raas}},
  \bibinfo{author}{\bibfnamefont{A.}~\bibnamefont{B\"uhler}}, \bibnamefont{and}
  \bibinfo{author}{\bibfnamefont{G.~S.} \bibnamefont{Uhrig}},
  \bibinfo{journal}{Eur. Phys. J. B} \textbf{\bibinfo{volume}{21}},
  \bibinfo{pages}{369} (\bibinfo{year}{2001}).

\bibitem[{\citenamefont{Sandvik et~al.}(1997)\citenamefont{Sandvik, Singh, and
  Campbell}}]{sandv97}
\bibinfo{author}{\bibfnamefont{A.~W.} \bibnamefont{Sandvik}},
  \bibinfo{author}{\bibfnamefont{R.~R.~P.} \bibnamefont{Singh}},
  \bibnamefont{and} \bibinfo{author}{\bibfnamefont{D.~K.}
  \bibnamefont{Campbell}}, \bibinfo{journal}{Phys. Rev. B}
  \textbf{\bibinfo{volume}{56}}, \bibinfo{pages}{14510} (\bibinfo{year}{1997}).

\bibitem[{\citenamefont{Sandvik and Campbell}(1999)}]{sandv99}
\bibinfo{author}{\bibfnamefont{A.~W.} \bibnamefont{Sandvik}} \bibnamefont{and}
  \bibinfo{author}{\bibfnamefont{D.~K.} \bibnamefont{Campbell}},
  \bibinfo{journal}{Phys. Rev. Lett.} \textbf{\bibinfo{volume}{83}},
  \bibinfo{pages}{195} (\bibinfo{year}{1999}).

\bibitem[{\citenamefont{K\"uhne and L\"ow}(1999)}]{kuehne99}
\bibinfo{author}{\bibfnamefont{R.~W.} \bibnamefont{K\"uhne}} \bibnamefont{and}
  \bibinfo{author}{\bibfnamefont{U.}~\bibnamefont{L\"ow}},
  \bibinfo{journal}{Phys. Rev. B} \textbf{\bibinfo{volume}{60}},
  \bibinfo{pages}{12125} (\bibinfo{year}{1999}).

\bibitem[{\citenamefont{Affleck et~al.}(1989)\citenamefont{Affleck, Gepner,
  Schulz, and Ziman}}]{affle89}
\bibinfo{author}{\bibfnamefont{I.}~\bibnamefont{Affleck}},
  \bibinfo{author}{\bibfnamefont{D.}~\bibnamefont{Gepner}},
  \bibinfo{author}{\bibfnamefont{H.~J.} \bibnamefont{Schulz}},
  \bibnamefont{and} \bibinfo{author}{\bibfnamefont{T.}~\bibnamefont{Ziman}},
  \bibinfo{journal}{J. Phys. A: Math. Gen.} \textbf{\bibinfo{volume}{22}},
  \bibinfo{pages}{511} (\bibinfo{year}{1989}).

\bibitem[{\citenamefont{Okamoto and Nomura}(1992)}]{okamo92}
\bibinfo{author}{\bibfnamefont{K.}~\bibnamefont{Okamoto}} \bibnamefont{and}
  \bibinfo{author}{\bibfnamefont{K.}~\bibnamefont{Nomura}},
  \bibinfo{journal}{Phys. Lett.} \textbf{\bibinfo{volume}{A169}},
  \bibinfo{pages}{433} (\bibinfo{year}{1992}).

\bibitem[{\citenamefont{Eggert}(1996)}]{egger96}
\bibinfo{author}{\bibfnamefont{S.}~\bibnamefont{Eggert}},
  \bibinfo{journal}{Phys. Rev. B} \textbf{\bibinfo{volume}{54}},
  \bibinfo{pages}{9612} (\bibinfo{year}{1996}).

\bibitem[{\citenamefont{Wegner}(1994)}]{wegne94}
\bibinfo{author}{\bibfnamefont{F.~J.} \bibnamefont{Wegner}},
  \bibinfo{journal}{Ann. Physik} \textbf{\bibinfo{volume}{3}},
  \bibinfo{pages}{77} (\bibinfo{year}{1994}).

\bibitem[{\citenamefont{Wegner}(2001)}]{wegne01a}
\bibinfo{author}{\bibfnamefont{F.~J.} \bibnamefont{Wegner}},
  \bibinfo{journal}{Phys. Rep.} \textbf{\bibinfo{volume}{348}},
  \bibinfo{pages}{77} (\bibinfo{year}{2001}).

\bibitem[{\citenamefont{Fr\"ohlich}(1952)}]{frohl52}
\bibinfo{author}{\bibfnamefont{H.}~\bibnamefont{Fr\"ohlich}},
  \bibinfo{journal}{Phys. Roy. Soc. Lond.} \textbf{\bibinfo{volume}{A215}},
  \bibinfo{pages}{291} (\bibinfo{year}{1952}).

\bibitem[{\citenamefont{Kehrein et~al.}(1996)\citenamefont{Kehrein, Mielke, and
  Neu}}]{kehre96a}
\bibinfo{author}{\bibfnamefont{S.~K.} \bibnamefont{Kehrein}},
  \bibinfo{author}{\bibfnamefont{A.}~\bibnamefont{Mielke}}, \bibnamefont{and}
  \bibinfo{author}{\bibfnamefont{P.}~\bibnamefont{Neu}}, \bibinfo{journal}{Z.
  Phys. B} \textbf{\bibinfo{volume}{99}}, \bibinfo{pages}{269}
  (\bibinfo{year}{1996}).

\bibitem[{\citenamefont{Mielke}(1998)}]{mielk98}
\bibinfo{author}{\bibfnamefont{A.}~\bibnamefont{Mielke}},
  \bibinfo{journal}{Eur. Phys. J. B} \textbf{\bibinfo{volume}{5}},
  \bibinfo{pages}{605} (\bibinfo{year}{1998}).

\bibitem[{\citenamefont{Knetter and Uhrig}(2000)}]{knett00a}
\bibinfo{author}{\bibfnamefont{C.}~\bibnamefont{Knetter}} \bibnamefont{and}
  \bibinfo{author}{\bibfnamefont{G.~S.} \bibnamefont{Uhrig}},
  \bibinfo{journal}{Eur. Phys. J. B} \textbf{\bibinfo{volume}{13}},
  \bibinfo{pages}{209} (\bibinfo{year}{2000}).

\bibitem[{\citenamefont{Fehske et~al.}(2000)\citenamefont{Fehske, Holicki, and
  Wei\ss{}e}}]{fehsk00}
\bibinfo{author}{\bibfnamefont{H.}~\bibnamefont{Fehske}},
  \bibinfo{author}{\bibfnamefont{M.}~\bibnamefont{Holicki}}, \bibnamefont{and}
  \bibinfo{author}{\bibfnamefont{A.}~\bibnamefont{Wei\ss{}e}}, in
  \emph{\bibinfo{booktitle}{Advances in Solid State Physics}}, edited by
  \bibinfo{editor}{\bibfnamefont{B.}~\bibnamefont{Kramer}}
  (\bibinfo{publisher}{Springer}, \bibinfo{address}{Berlin},
  \bibinfo{year}{2000}), vol.~\bibinfo{volume}{40}, p. \bibinfo{pages}{235}.

\bibitem[{\citenamefont{Geertsma and Khomskii}(1996)}]{geert96}
\bibinfo{author}{\bibfnamefont{W.}~\bibnamefont{Geertsma}} \bibnamefont{and}
  \bibinfo{author}{\bibfnamefont{D.}~\bibnamefont{Khomskii}},
  \bibinfo{journal}{Phys. Rev. B} \textbf{\bibinfo{volume}{54}},
  \bibinfo{pages}{3011} (\bibinfo{year}{1996}).

\bibitem[{\citenamefont{Khomskii et~al.}(1996)\citenamefont{Khomskii, Geertsma,
  and Mostovoy}}]{khoms96}
\bibinfo{author}{\bibfnamefont{D.}~\bibnamefont{Khomskii}},
  \bibinfo{author}{\bibfnamefont{W.}~\bibnamefont{Geertsma}}, \bibnamefont{and}
  \bibinfo{author}{\bibfnamefont{M.}~\bibnamefont{Mostovoy}},
  \bibinfo{journal}{Czech. Journ. of Physics} \textbf{\bibinfo{volume}{46}},
  \bibinfo{pages}{3239} (\bibinfo{year}{1996}).

\bibitem[{\citenamefont{Werner et~al.}(1999)\citenamefont{Werner, Gros, and
  Braden}}]{werne99}
\bibinfo{author}{\bibfnamefont{R.}~\bibnamefont{Werner}},
  \bibinfo{author}{\bibfnamefont{C.}~\bibnamefont{Gros}}, \bibnamefont{and}
  \bibinfo{author}{\bibfnamefont{M.}~\bibnamefont{Braden}},
  \bibinfo{journal}{Phys. Rev. B} \textbf{\bibinfo{volume}{59}},
  \bibinfo{pages}{14356} (\bibinfo{year}{1999}).

\bibitem[{\citenamefont{B\"uhler et~al.}(2000)\citenamefont{B\"uhler, Elstner,
  and Uhrig}}]{buehler00a}
\bibinfo{author}{\bibfnamefont{A.}~\bibnamefont{B\"uhler}},
  \bibinfo{author}{\bibfnamefont{N.}~\bibnamefont{Elstner}}, \bibnamefont{and}
  \bibinfo{author}{\bibfnamefont{G.~S.} \bibnamefont{Uhrig}},
  \bibinfo{journal}{Eur. Phys. J. B} \textbf{\bibinfo{volume}{16}},
  \bibinfo{pages}{475} (\bibinfo{year}{2000}).

\bibitem[{\citenamefont{Gros and Werner}(1998)}]{gros98}
\bibinfo{author}{\bibfnamefont{C.}~\bibnamefont{Gros}} \bibnamefont{and}
  \bibinfo{author}{\bibfnamefont{R.}~\bibnamefont{Werner}},
  \bibinfo{journal}{Phys. Rev. B} \textbf{\bibinfo{volume}{58}},
  \bibinfo{pages}{14677} (\bibinfo{year}{1998}).

\bibitem[{\citenamefont{Uhrig and Normand}(2001)}]{uhrig00a}
\bibinfo{author}{\bibfnamefont{G.~S.} \bibnamefont{Uhrig}} \bibnamefont{and}
  \bibinfo{author}{\bibfnamefont{B.}~\bibnamefont{Normand}},
  \bibinfo{journal}{Phys. Rev. B} \textbf{\bibinfo{volume}{63}},
  \bibinfo{pages}{134418} (\bibinfo{year}{2001}).

\bibitem[{\citenamefont{B\"uhler et~al.}(2001)\citenamefont{B\"uhler, L\"ow,
  and Uhrig}}]{buehler01a}
\bibinfo{author}{\bibfnamefont{A.}~\bibnamefont{B\"uhler}},
  \bibinfo{author}{\bibfnamefont{U.}~\bibnamefont{L\"ow}}, \bibnamefont{and}
  \bibinfo{author}{\bibfnamefont{G.~S.} \bibnamefont{Uhrig}},
  \bibinfo{journal}{Phys. Rev. B} \textbf{\bibinfo{volume}{64}},
  \bibinfo{pages}{024428} (\bibinfo{year}{2001}).

\bibitem[{\citenamefont{Kl\"umper}(1993)}]{klump93b}
\bibinfo{author}{\bibfnamefont{A.}~\bibnamefont{Kl\"umper}},
  \bibinfo{journal}{Z. Phys. B} \textbf{\bibinfo{volume}{91}},
  \bibinfo{pages}{507} (\bibinfo{year}{1993}).

\bibitem[{\citenamefont{Eggert et~al.}(1994)\citenamefont{Eggert, Affleck, and
  Takahashi}}]{egger94}
\bibinfo{author}{\bibfnamefont{S.}~\bibnamefont{Eggert}},
  \bibinfo{author}{\bibfnamefont{I.}~\bibnamefont{Affleck}}, \bibnamefont{and}
  \bibinfo{author}{\bibfnamefont{M.}~\bibnamefont{Takahashi}},
  \bibinfo{journal}{Phys. Rev. Lett.} \textbf{\bibinfo{volume}{73}},
  \bibinfo{pages}{332} (\bibinfo{year}{1994}).

\bibitem[{\citenamefont{Kl\"umper}(1998)}]{klump98a}
\bibinfo{author}{\bibfnamefont{A.}~\bibnamefont{Kl\"umper}},
  \bibinfo{journal}{Eur. Phys. J. B} \textbf{\bibinfo{volume}{5}},
  \bibinfo{pages}{677} (\bibinfo{year}{1998}).

\bibitem[{\citenamefont{Evertz et~al.}(1993)\citenamefont{Evertz, Lana, and
  Marcu}}]{evertz93}
\bibinfo{author}{\bibfnamefont{H.~G.} \bibnamefont{Evertz}},
  \bibinfo{author}{\bibfnamefont{G.}~\bibnamefont{Lana}}, \bibnamefont{and}
  \bibinfo{author}{\bibfnamefont{M.}~\bibnamefont{Marcu}},
  \bibinfo{journal}{Phys. Rev. Lett.} \textbf{\bibinfo{volume}{70}},
  \bibinfo{pages}{875} (\bibinfo{year}{1993}).

\bibitem[{\citenamefont{Castilla et~al.}(1995)\citenamefont{Castilla,
  Chakravarty, and Emery}}]{casti95}
\bibinfo{author}{\bibfnamefont{G.}~\bibnamefont{Castilla}},
  \bibinfo{author}{\bibfnamefont{S.}~\bibnamefont{Chakravarty}},
  \bibnamefont{and} \bibinfo{author}{\bibfnamefont{V.~J.} \bibnamefont{Emery}},
  \bibinfo{journal}{Phys. Rev. Lett.} \textbf{\bibinfo{volume}{75}},
  \bibinfo{pages}{1823} (\bibinfo{year}{1995}).

\bibitem[{\citenamefont{Riera and Dobry}(1995)}]{riera95}
\bibinfo{author}{\bibfnamefont{J.}~\bibnamefont{Riera}} \bibnamefont{and}
  \bibinfo{author}{\bibfnamefont{A.}~\bibnamefont{Dobry}},
  \bibinfo{journal}{Phys. Rev. B} \textbf{\bibinfo{volume}{51}},
  \bibinfo{pages}{16098} (\bibinfo{year}{1995}).

\bibitem[{\citenamefont{Fabricius et~al.}(1998)\citenamefont{Fabricius,
  Kl\"umper, L\"ow, B\"uchner, Lorenz, Dhalenne, and Revcolevschi}}]{fabri98a}
\bibinfo{author}{\bibfnamefont{K.}~\bibnamefont{Fabricius}},
  \bibinfo{author}{\bibfnamefont{A.}~\bibnamefont{Kl\"umper}},
  \bibinfo{author}{\bibfnamefont{U.}~\bibnamefont{L\"ow}},
  \bibinfo{author}{\bibfnamefont{B.}~\bibnamefont{B\"uchner}},
  \bibinfo{author}{\bibfnamefont{T.}~\bibnamefont{Lorenz}},
  \bibinfo{author}{\bibfnamefont{G.}~\bibnamefont{Dhalenne}}, \bibnamefont{and}
  \bibinfo{author}{\bibfnamefont{A.}~\bibnamefont{Revcolevschi}},
  \bibinfo{journal}{Phys. Rev. B} \textbf{\bibinfo{volume}{57}},
  \bibinfo{pages}{1102} (\bibinfo{year}{1998}).

\end{thebibliography}

\end{document}